\documentclass[doublecol]{epl2}
\usepackage{amsmath}
\usepackage{color}
\usepackage{hyperref}
\usepackage{graphicx}

\title{Quantum Oscillation as Diagnostics of Pseudogap State in Underdoped Cuprates}
\author{Long Zhang\inst{1,2} \and Jia-Wei Mei\inst{3}}
\shortauthor{L. Zhang \and J.-W. Mei}
\institute{                    
  \inst{1} International Center for Quantum Materials, School of Physics, Peking University, Beijing, 100871, China\\
  \inst{2} Institute for Advanced Study, Tsinghua University, Beijing, 100084, China\\
  \inst{3} Perimeter Institute for Theoretical Physics, Waterloo, Ontario, N2L 2Y5, Canada
}

\abstract{
The Fermi surface in underdoped cuprates is reconstructed by the charge density wave (CDW) order in the pseudogap phase. Theoretical proposals can be divided into two classes: one assumes the underlying Fermi surface without CDW as a conventional large surface; the other assumes small hole-like Fermi pockets. In both scenarios, we theoretically study the quantum oscillation and find three evenly spaced peaks in the oscillation spectra. The central dominant peak is induced by the CDW order. Its effective mass is strongly enhanced as the CDW vanishes in agreement with experiments. But the two scenarios have different understandings of the subdominant satellite peaks. In the large-surface scenario they are induced by the interlayer tunneling between the bilayer CuO$_{2}$ planes. Their effective masses are also enhanced with descreasing CDW. In the small-pocket scenario one of the subdominant peaks comes from the original small Fermi pockets of the pseudogap state. Its effective mass is nearly independent of the CDW strength and increases monotonically with the doping. We propose future quantum oscillation experiments to test these different predictions and thus to clarify the underlying Fermi surface structure of the pseudogap state.
}
\begin{document}
\maketitle

\section{Introduction}
The nature of the pseudogap phase is the holy grail in the study of cuprate superconductors \cite{Timusk1999, Keimer2015}. The observations of quantum oscillation (QO) and charge density wave (CDW) in the pseudogap phase in underdoped YBa$_{2}$Cu$_{3}$O$_{6+\delta}$ (YBCO) \cite{Doiron-Leyraud2007, Wu2011b, Ghiringhelli2012, Chang2012} and HgBa$_{2}$CuO$_{4+\delta}$ (Hg1201) \cite{Barisic2013, Tabis2014a} brought much excitement in the community. Both QO and long-range CDW order show up in the same doping range $0.08<x<0.16$ at low temperature when the superconductivity is suppressed in high magnetic fields \cite{Sebastian2010, Ramshaw2014, Blanco-Canosa2014, Huecker2014, Wu2011b, LeBoeuf2012}. In the same regime the Hall and Seebeck coefficients change sign from positive to negative \cite{LeBoeuf2007, LeBoeuf2011, Laliberte2011, Doiron-Leyraud2013, Badoux2015} showing that the d.c. charge transport is dominated by an electron-like Fermi pocket.

\begin{figure}
\centering
\includegraphics[width=0.45\textwidth]{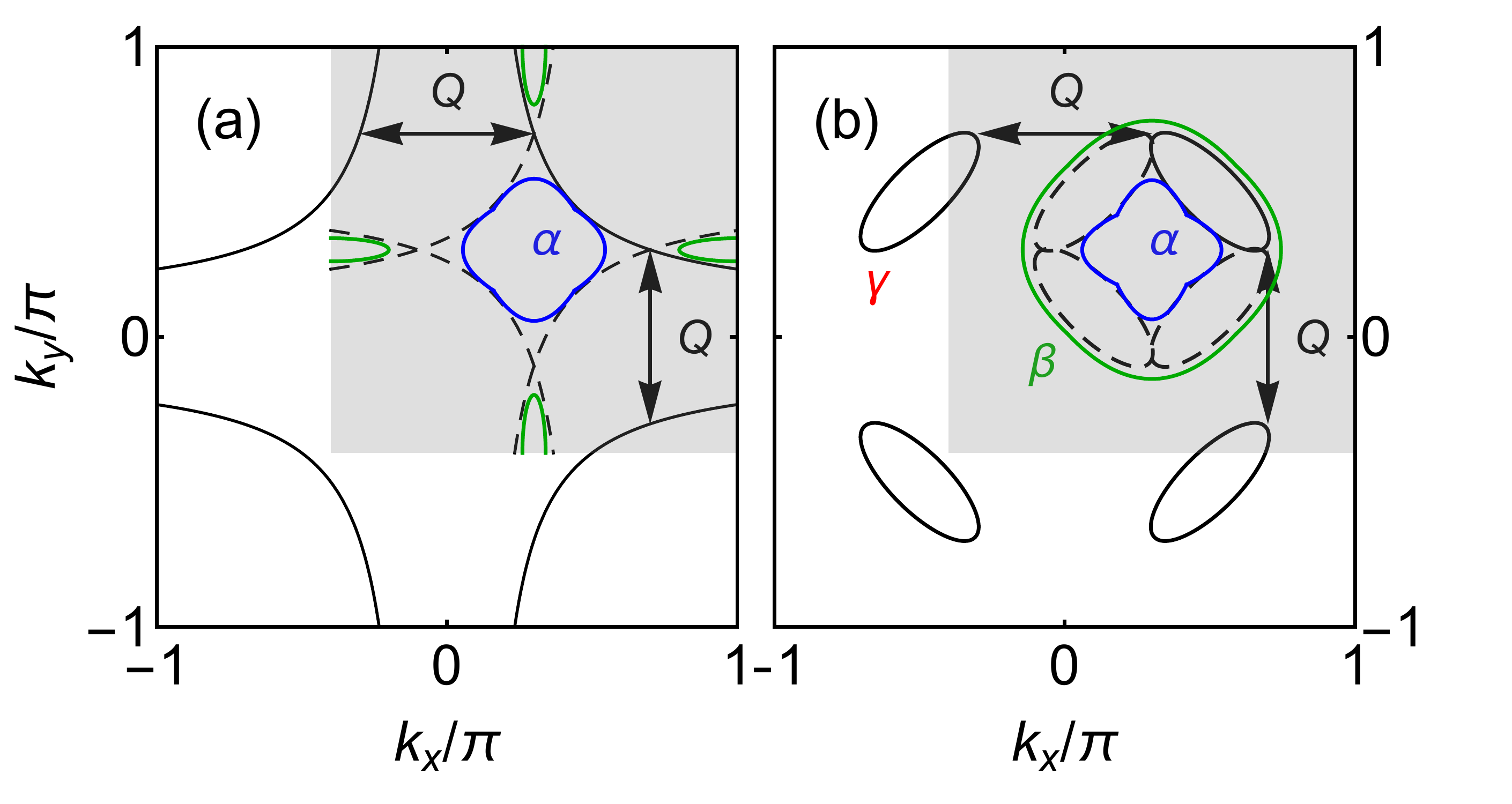}
\caption{(a) Large-surface and (b) small-pocket scenarios of the CDW-induced Fermi surface reconstruction. The arrows indicate the biaxial CDW wavevectors $(Q,0)$ and $(0,Q)$. In the folded Brillouin zone (shaded region) the original Fermi surfaces (black) are reconstructed into new small (blue) and large (green) Fermi pockets.}
\label{FigPhase}
\end{figure}

It is widely accepted that the QO and the transport anomalies ensue from the CDW order-induced Fermi surface reconstruction \cite{Sebastian2015}, but the nature of the ``normal'' state in the absence of CDW order remains elusive. Theoretical proposals are roughly divided into two classes. One assumes that without CDW order the state is a conventional metal with a large hole-like Fermi surface (``large-surface'' scenario) \cite{Allais2014}. CDW order breaks up the large Fermi surface into small pockets as illustrated in Fig. \ref{FigPhase} (a). The CDW fluctuations lead to pseudogap phenomena, e.g., the unclosed arc-like Fermi surface observed in angle-resolved photoemission spectroscopy (ARPES) \cite{Harrison2014}.

In contrast, other theoretical proposals assume that the pseudogap phenomena originate in different mechanisms \cite{Lee2006, Yang2006, Baskaran2007, Senthil2009, Qi2010a, Moon2011, Rice2012, Mei2012a, Vojta2012, Bejas2012, Ma2013, Efetov2013, Lee2014}, e.g., strong Coulomb repulsion or antiferromagnetic fluctuations. The metallic pseudogap state has only small hole-like Fermi pockets (``small-pocket'' scenario) \cite{Yang2006, Baskaran2007, Qi2010a, Moon2011, Mei2012a, Ma2013}. It is called a Luttinger-volume violating Fermi liquid in Ref. \cite{Mei2012a}. This is  because in a conventional metal the area of Fermi surfaces (spin degeneracy counted) amounts to $(1+x)S_{\mathrm{BZ}}$ (Luttinger theorem) while in the small-pocket scenario it amounts to $x S_{\mathrm{BZ}}$ instead. Here $x$ is the hole doping concentration away from the half-filled compound and $S_{\mathrm{BZ}}$ is the area of the first Brillouin zone. This leads to the suppression of low-energy density of states (DoS) in the pseudogap phase \cite{Lee2006, Rice2012}. The CDW order is taken as a secondary instability of the pseudogap state \cite{Zhang2014b, Chowdhury2014, Atkinson2015}. It reconstructs the Fermi surface as illustrated in Fig. \ref{FigPhase} (b).

In this work we will compare these two different scenarios by studying the QO spectrum for the CDW reconstructed Fermi surface. In both scenarios the QO is dominated by the CDW-induced small electron pocket (denoted by $\alpha$ in Fig. \ref{FigPhase}). In experiments, the $\alpha$ peak and two subdominant peaks form an evenly spaced \emph{three-peak} structure \cite{Sebastian2010b} (the left one is denoted by $\gamma$ and the right one by $\delta$, respectively). The two scenarios have very different understandings of the two subdominant peaks. In the large-surface scenario, one has to resort to the CuO$_{2}$ bilayer structure of YBCO. The multiple peaks develop due to the interlayer tunneling and the induced magnetic breakdown orbits \cite{Audouard2009, Garcia-Aldea2010, Harrison2011, Sebastian2012a, Maharaj2015}. On the other hand, in the small-pocket scenario we will show that the bilayer structure is not necessary to explain the multiple peaks. The subdominant $\gamma$ peak comes from the original small hole pockets of the pseudogap state while the $\delta$ peak comes from the $2\alpha-\gamma$ orbit due to the magnetic breakdown mechanism.

The different understandings of the three-peak structure imply distinct predictions, which can be tested with the QO experiments. In the large-surface scenario, since all multiple peaks are induced by the CDW order, all of their effective masses are strongly enhanced by descreasing the CDW order. In the small-pocket scenario, however, the effective mass enhancement is only found for the CDW-induced $\alpha$ pocket. The $\gamma$ pocket effective mass $m_{\gamma}^{*}$ is not sensitive to the CDW strength and increases monotonically with the doping.

In cuprates the CDW order vanishes as the doping approaches the critical points $x_{c1}\simeq 0.08$ and $x_{c2}\simeq 0.16$. The effective mass of the dominant peak $m_{\alpha}^{*}$ is strongly enhanced \cite{Sebastian2010, Ramshaw2014} consistent with both scenarios studied in this work, but the evolution of the effective mass of the subdominant $\gamma$ peak $m_{\gamma}^{*}$ has not been systematically examined to date. We propose that measuring the evolution of $m_{\gamma}^{*}$ with doping is a feasible way to judge these two theoretical scenarios and thus to clarify the nature of the pseudogap phase.

\section{Method}
The CDW order is modeled with the following effective CDW Hamiltonian,
\begin{equation} \label{EqCDW}
H_{\mathrm{CDW}}= \sum_{\vec{k},\sigma}f(\vec{k})\sum_{i=1,2}c_{\vec{k}+\vec{Q}_{i}/2,\sigma}^{\dag}c_{\vec{k}-\vec{Q}_{i}/2,\sigma}.
\end{equation}
In accord with experiments in YBCO \cite{Fujita2014, Comin2014, Achkar2014} the CDW order is taken to be of $d$ form: $f(\vec{k})=\cos k_{x}-\cos k_{y}$ \cite{Sachdev2013a}. We note that different choices of the form factor do not change our main results. The biaxial CDW order with wavevectors $\vec{Q}_{1}=(Q, 0)$ and $\vec{Q}_{2}=(0,Q)$ are implemented to reconstruct the Fermi surface in both scenarios as illustrated in Fig. \ref{FigPhase}.

The quantum oscillation is characterized by the DoS at the Fermi level in magnetic fields, $D_{\eta}(B)=-\pi^{-1}\mathrm{Im}\mathrm{Tr}G(i\eta,\vec{k}-e\vec{A}^{e})$, in which $G$ is the electron Green's function for the full Hamiltonian $H=H_0+P_0H_\text{CDW}$. The model Hamiltonians $H_0$ for different scenarios are given in Eq. (\ref{EqLarge}) and (\ref{EqH0}), respectively. $P_0$ is the CDW strenghth. $\eta$ is a Lorentz broadening parameter. $\vec{A}^{e}$ is the magnetic vector potential. We calculate $D_{\eta}(B)$ on a large lattice with more than $3\times 10^{4}$ sites to avoid finite size effect and take Fourier transformation to obtain the QO frequency spectrum. The QO frequency $F$ is related to the Fermi pocket area $S$ by the Onsager relation \cite{Shoenberg1984magnetic}, $F=\hbar S/2\pi e$.

In QO experiments the effective masses are extracted from the temperature dependence of the oscillation amplitudes by fitting the Lifshitz-Kosevich formula \cite{Shoenberg1984magnetic}. In our theoretical calculations the Lorentz broadening $\eta$ in $D_{\eta}(B)$ serves as an effective temperature in smearing the DoS oscillation. The oscillation amplitude $R_{\eta}$ decreases exponentially with $\eta$: $R_{\eta}=e^{-2\pi m^{*}\eta/e\hbar B}$ \cite{Shoenberg1984magnetic}, so we can deduce the effective mass of each orbit from the decay rate of $R_{\eta}$.

\begin{figure}
\centering
\includegraphics[width=0.45\textwidth]{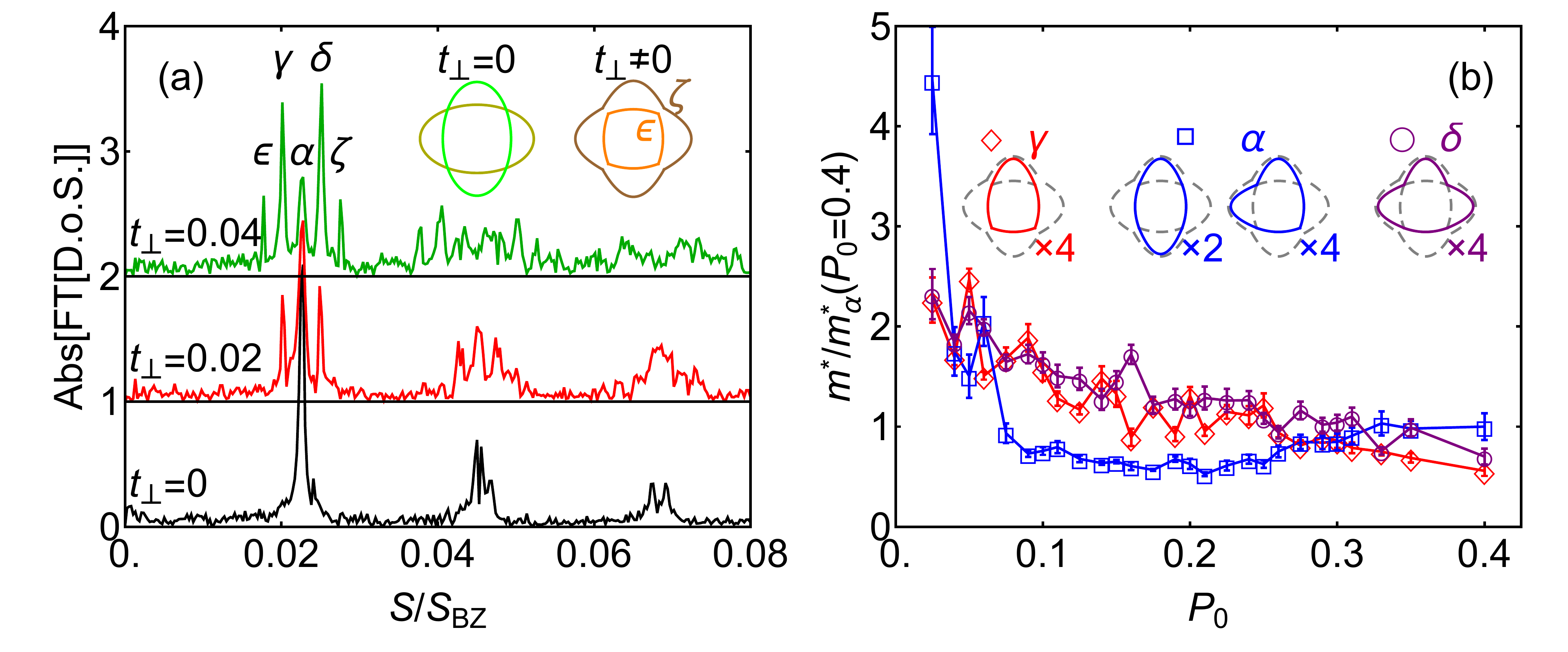}
\caption{DoS oscillation in the bilayer splitting model. (a) Fourier spectra of the DoS oscillation for different $t_{\perp}$ (offset for clarity). Inset: Schematic illustration of the CDW-induced bilayer Fermi surfaces with or without interlayer tunneling. (b) Effective masses $m_{\alpha, \gamma, \delta}^{*}$ vs. CDW strength $P_{0}$ for $t_{\perp}=0.02$. These magnetic breakdown orbits are illustrated in the inset. Parameters: $t=1$, $t'=-0.2$, $\mu=-0.7$, $t_{\delta}=0.15$. The doping $x=0.13$. The CDW wavevector $Q$ is taken to be $2\pi /3$ for simplicity. $P_{0}=0.3$ in Panel (a). Error bars in (b) come from fitting $R_{\eta}$.}
\label{FigBilayer}
\end{figure}

\section{QO in CDW-reconstructed large-surface model}
In the large-surface scenario, a biaxial CDW order reconstructs the large hole-like Fermi surface into small pockets as sketched in Fig. \ref{FigPhase} (a). To account for the experimental multiple frequency QO spectrum, we adopt the following bilayer model given by $H=H_{0}^{\mathrm{b.l.}}+P_{0}H_{\mathrm{CDW}}$, where
\begin{eqnarray}
\label{EqLarge}
H_{0}^{\mathrm{b.l.}}=\sum_{\vec{k},\sigma}c_{\vec{k},\sigma}^{\dagger}h(\vec{k})c_{\vec{k},\sigma},
\end{eqnarray}
with $h(\vec{k}) = -2t(\cos k_{x}+\cos k_{y}) - 4t'\cos k_{x}\cos k_{y} - \mu - t_{\perp}\tau_{x} - 2t_{\delta}(\cos k_{x}-\cos k_{y})\tau_{z}$. $\tau_{x,z}$ are the Pauli matrices acting on the bilayer indices. The $t_{\delta}$ term introduces an inplane anisotropy and breaks the bilayer mirror symmetry, which is necessary for the emergence of magnetic breakdown orbits \cite{Maharaj2015}. As shown in Fig. \ref{FigBilayer} insets, the CDW-induced diamond-shaped Fermi pockets are elongated by the inplane anisotropy along the $k_{x}$ and $k_{y}$ directions in the two layers, respectively. The interlayer tunneling $t_{\perp}$ splits them into two Fermi surfaces and induces another four magnetic breakdown orbits in magnetic fields. The small hole pockets sketched in Fig. \ref{FigPhase} (a) (green) and open Fermi sheets may appear at weak CDW order as discussed in Ref. \cite{Allais2014}, but are suppressed by strong CDW order (but not too strong to destroy the QO \cite{Zhang2015}).

Making the Peierls substitution $t_{ij}\mapsto t_{ij}e^{-ieA_{ij}^{e}}$ on a lattice, we calculate the DoS $D_{\eta}(B)$ in magnetic fields and take Fourier transformation for the QO frequency spectra, which are shown in Fig. \ref{FigBilayer} (a). Those orbits from the bilayer splitting and magnetic breakdown give rise to the multiple QO peaks. The three-peak pattern is found for small but nonzero $t_{\perp}$ in particular. The effective masses of the three peaks are extracted from the $\eta$ dependence of the peak heights at different CDW strength $P_{0}$, which are shown in Fig. \ref{FigBilayer} (b). When $P_{0}$ vanishes, the effective masses of all these peaks are strongly enhanced.

\begin{figure}
\centering
\includegraphics[width=0.45\textwidth]{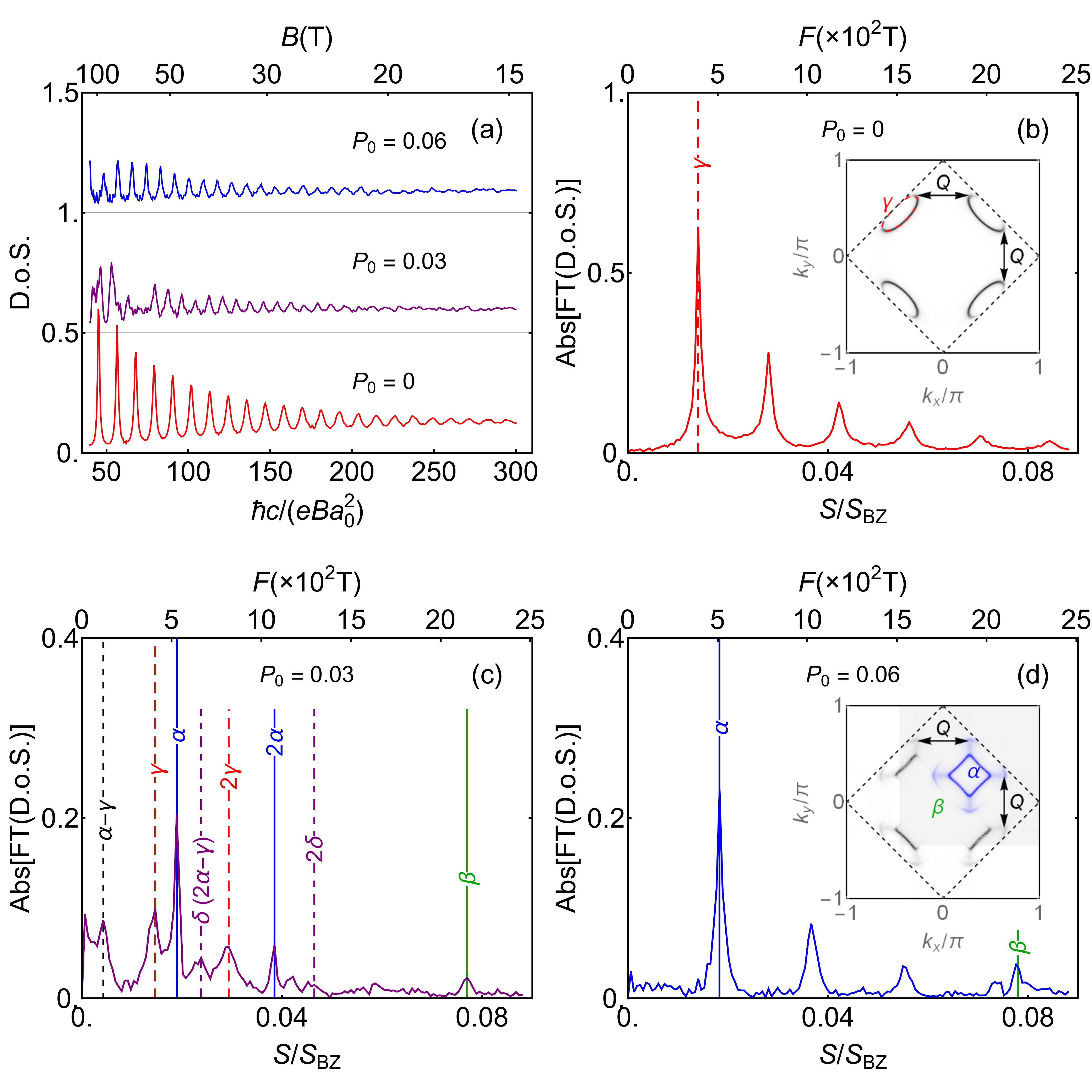}
\caption{(a) DoS oscillation in the YRZ model for different CDW strengths, $P_{0}=0$, $0.03$, and $0.06$ (offset for clarity). $x=0.12$. (b)--(d) The Fourier transformations in which the multiple oscillation peaks are labeled. Insets: The corresponding spectral functions at the Fermi level.}
\label{FigDoSQO}
\end{figure}

\section{QO in CDW-reconstructed small-pocket model}
In the small-pocket scenario, the metallic pseudogap state violates the conventional Luttinger theorem and has only small hole-like Fermi pockets. Several models have been proposed to describe the small hole pockets in the pseudogap state \cite{Yang2006, Qi2010a, Moon2011, Mei2012a, Ma2013}. We will adopt the Yang-Rice-Zhang (YRZ) phenomenological Green's function \cite{Yang2006} to describe the pseudogap state for its simplicity and popularity,
\begin{equation} \label{EqYRZ}
G_{0}(\omega, \vec{k})= \frac{g_{t}(x)}{\omega-\xi(\vec{k})-\Sigma_{\mathrm{RVB}}(\omega, \vec{k})},
\end{equation}
in which $\xi(\vec{k})=-2t(x)(\cos k_{x}+\cos k_{y})-4t'(x)\cos k_{x}\cos k_{y}-2t''(x)(\cos 2k_{x}+\cos 2k_{y})-\mu(x)$. The self-energy $\Sigma_{\mathrm{RVB}}(\omega, \vec{k}) = \Delta(\vec{k})^{2}/(\omega +\xi_{0}(\vec{k}))$ with $\xi_{0}(\vec{k})=-2t(x)(\cos k_{x}+\cos k_{y})$ and $\Delta(\vec{k})=\Delta_{0}(x)(\cos k_{x}-\cos k_{y})$. The hopping parameters $t(x)=g_{t}(x)t+3g_{J}(x)J\chi(x)/8$, $t'(x)=g_{t}(x)t'$ and $t''(x)=g_{t}(x)t''$ are renormalized from the bare band structure parameters $t$, $t'=-0.3t$, $t''=0.2t$ and $J=t/3$. $g_{t}(x)=2x/(1+x)$ and $g_{J}(x)=4/(1+x)^{2}$ capture the impact of the strong onsite Coulomb repulsion. $\chi(x)$ and $\Delta_{0}(x)$ are determined self-consistently with the renormalized mean field theory of the $t$-$J$ model \cite{Zhang1988}. This model can be derived based on the slave-boson theory of doped Mott insulators \cite{James2012}.

The YRZ model has four hole-like Fermi pockets (denoted by $\gamma$) in the nodal region, i.e., near $(\pi/2,\pi/2)$ as shown in Fig. \ref{FigDoSQO} (b) inset. The chemical potential $\mu(x)$ is adjusted so that the total area of hole pockets (spin degeneracy counted) amounts to $xS_{\mathrm{BZ}}$, so the area of each pocket $S_{\gamma}$ satisfies
\begin{equation} \label{EqSgamma}
S_{\gamma}/S_{\mathrm{BZ}}= x/8.
\end{equation}
It is successful in describing a range of pseudogap phenomena in cuprates \cite{Rice2012}.

In order to calculate the DoS in magnetic fields we introduce the following two-band effective Hamiltonian \cite{Qi2010a},
\begin{equation} \label{EqH0}
H_{0}=\sum_{\vec{k},\sigma}
\begin{pmatrix}
c_{\sigma\vec{k}}^{\dag}&\tilde{c}_{\sigma\vec{k}}^{\dag}
\end{pmatrix}
\begin{pmatrix}
\xi(\vec{k})& \Delta(\vec{k})\\
\Delta(\vec{k})& -\xi_{0}(\vec{k})
\end{pmatrix}
\begin{pmatrix}
c_{\sigma\vec{k}}\\
\tilde{c}_{\sigma\vec{k}}
\end{pmatrix}.
\end{equation}
The YRZ Green's function in Eq. (\ref{EqYRZ}) can be rewritten as $G_{0}(\omega,\vec{k})=g_{t}(x)\langle \mathcal{T}_{t}c_{\sigma}(\omega,\vec{k})c_{\sigma}^{\dagger}(\omega,\vec{k})\rangle$ according to the effective Hamiltonian $H_{0}$ in Eq. (\ref{EqH0}) \cite{Qi2010a}. The biaxial CDW wavevectors connect the hotspots, i.e., the tips of the arc-like Fermi surfaces, and reconstructs the nodal $\gamma$ pockets into an $\alpha$ orbit and a large $\beta$ orbit [see Fig. \ref{FigDoSQO} (d) inset]. In magnetic fields, making the Peierls substitution in the Hamiltonian $H= H_{0}+P_{0}H_{\mathrm{CDW}}$, the DoS $D_{\eta}(B)$ is calculated and shown in Fig. \ref{FigDoSQO}. 

As the CDW strength $P_{0}$ increases, the QO peak from the original $\gamma$ pocket is gradually suppressed and gives way to the CDW-induced $\alpha$ and $\beta$ pockets. In addition to the $\alpha$ and $\beta$ pocket peaks, the $\gamma$ pocket peak also survives in moderate CDW order. Their coexistence generates the multiple QO peaks as labeled in Fig. \ref{FigDoSQO} (c). The $\gamma$ peak lies on the left side of the dominant $\alpha$ peak while the magnetic breakdown $2\alpha-\gamma$ orbit peak (denoted by $\delta$) lies on the right. They form the three-peak structure observed in experiments. We stress that in this scenario the three-peak pattern has nothing to do with the bilayer structure of YBCO. In experiments the three-peak structure is also discernible in the \emph{single-layer} Hg1201 compound \cite{Barisic2013}. The $\beta$ pocket corresponds to the large frequency peak found in Refs. \cite{Sebastian2008, Sebastian2010, Sebastian2010b, Sebastian2011}. The small-frequency peak observed in Ref. \cite{Doiron-Leyraud2014} comes from the $\alpha-\gamma$ orbit, which is also induced by magnetic breakdown. These peaks with particular relevance to experiments are summarized in Table \ref{TabMulti}.

\begin{table}
\centering
\caption{Correspondence of the multiple QO peaks found in experiments and in the small-pocket scenario. $S_{\mathrm{BZ}}$ is set to be unity.}
\label{TabMulti}
\begin{tabular}{cccc}
\hline \hline
Peak & Area	& Remarks & References \\
\hline
$\alpha$ & $S_{\alpha}$ & Main peak & \cite{Doiron-Leyraud2007, Barisic2013}\\
$\gamma$ & $S_{\gamma}=x/8$ & Left of triplet & \cite{Audouard2009, Singleton2010, Sebastian2010b, Barisic2013}\\
$\delta$ & $2S_{\alpha}-S_{\gamma}$ & Right of triplet & \cite{Audouard2009, Sebastian2010b, Barisic2013}\\
$\beta$	 & $S_{\alpha}+x/2$ & & \cite{Sebastian2008, Sebastian2010, Sebastian2010b, Sebastian2011}\\
$\alpha-\gamma$  & $S_{\alpha}-S_{\gamma}$ & & \cite{Doiron-Leyraud2014}\\
\hline \hline
\end{tabular}
\end{table}

\begin{figure}
\centering
\includegraphics[width=0.45\textwidth]{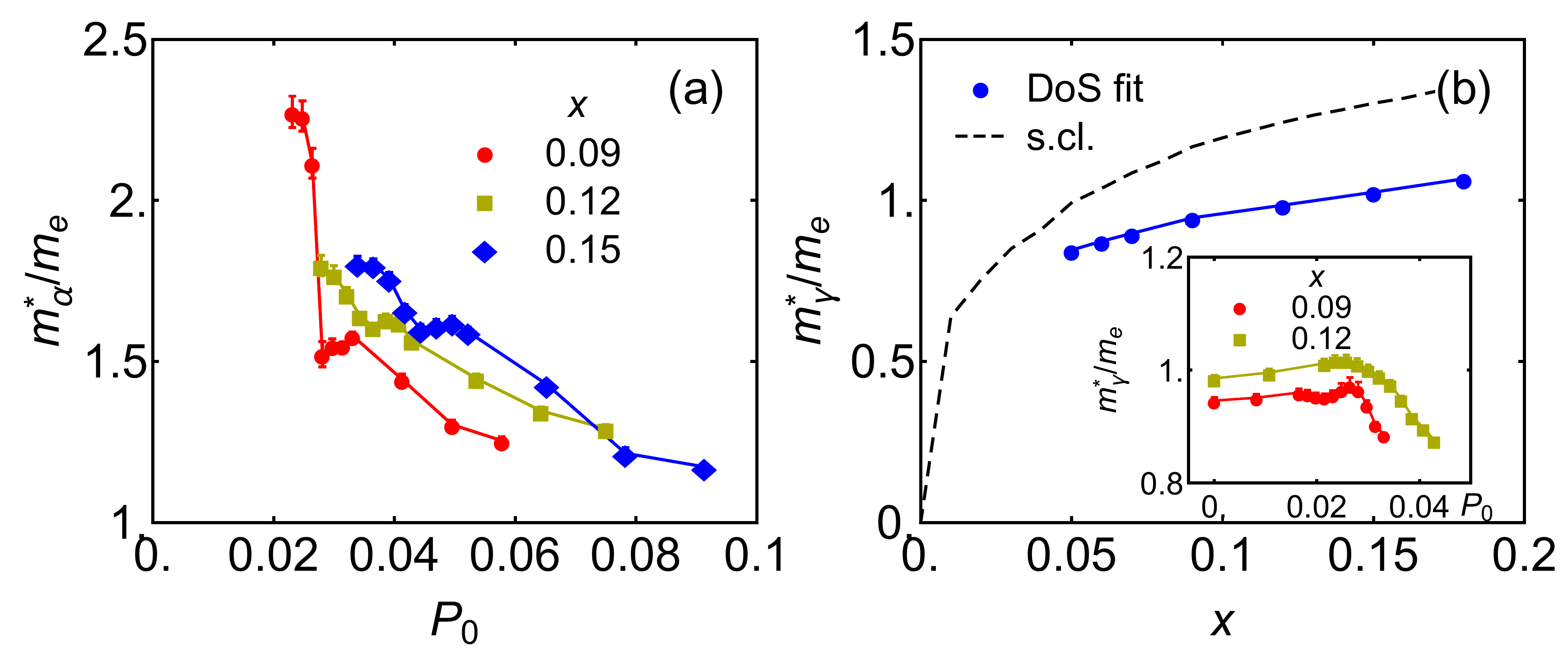}
\caption{(a) $m_{\alpha}^{*}$ enhancement with decreasing CDW order $P_{0}$. (b) Monotonic increase of $m_{\gamma}^{*}$ with doping in the absence of CDW order. Inset: $m_{\gamma}^{*}$ versus $P_{0}$. $m_{\gamma}^{*}$ is nearly constant as $P_{0}\rightarrow 0$. We take $t = 300~\mathrm{meV}$.}
\label{FigMasses}
\end{figure}

Their effective masses $m_{\alpha,\gamma}^{*}$ are shown in Fig. \ref{FigMasses}. When the CDW strength $P_{0}$ decreases, $m_{\alpha}^{*}$ is strongly enhanced [Fig. \ref{FigMasses} (a)], but $m_{\gamma}^{*}$ is nearly independent of $P_{0}$ for weak CDW order [Fig. \ref{FigMasses} (b) inset]. If the CDW order is absent or weak, $m_{\gamma}^{*}$ increases monotonically with doping, which is in sharp contrast to the nonmonotonic evolution of $m_{\alpha}^{*}$ in YBCO \cite{Sebastian2010, Ramshaw2014}. The semiclassical (s.cl.) result $m^{*}=\frac{\hbar^{2}}{2\pi}\frac{\partial S}{\partial \mu}$ is shown along for comparison. The monotonicity of $m_{\gamma}^{*}$ agrees with the optical Hall angle measurements in the pseudogap phase \cite{Rigal2004, Shi2005, Drew2007}.

\section{Discussion}
The $m_{\alpha}^{*}$ enhancement near the CDW critical points is qualitatively captured by both scenarios in agreement with experiments \cite{Sebastian2010, Ramshaw2014}. Theoretically the CDW dynamical fluctuations near the critical points can further enhance $m_{\alpha}^{*}$ \cite{Senthil2014}, but it is beyond our mean-field-type treatment of the static CDW order.

The different trends of $m_{\gamma}^{*}$ with the doping in these two scenarios are the key findings in this work. These two widely adopted scenarios represent very different perspectives of the pseudogap phase, but they have not been directly contrasted in the same set of experiments before. Measuring the trend of $m_{\gamma}^{*}$ in QO experiments provides an opportunity to compare the two scenarios and thus helps to clarity the nature of the pseudogap state.

In the large-surface scenario, the pseudogap phenomena are suggested to be driven by the CDW dynamical fluctuations \cite{Harrison2014}; however, experiments hint that the pseudogap and the CDW are of different origins despite that a general consensus has not been reached. First the biaxial CDW order found in YBCO and Hg1201 is not universal in the cuprate families. The La-based cuprates have a distinct stripe order for $x\simeq 1/8$ \cite{Tranquada1995}. But the pseudogap phenomena are quite universal in all hole-underdoped cuprates \cite{Timusk1999}. Second the doping and temperature range of the CDW order or fluctuations are much more restricted than the pseudogap \cite{Blanco-Canosa2014, Huecker2014}. Last but not least the charge carrier concentration measured with Hall resistivity jumps from $x$ to $1+x$ around the critical doping $x_{\mathrm{PG}}\simeq 0.19$ where the pseudogap phenomena vanish \cite{Badoux2015}. It suggests that the pseudogap state at $x < x_{\mathrm{PG}}$ has small hole pockets in contrast to the large Fermi surface at $x > x_{\mathrm{PG}}$.

\begin{figure}
\centering
\includegraphics[width=0.45\textwidth]{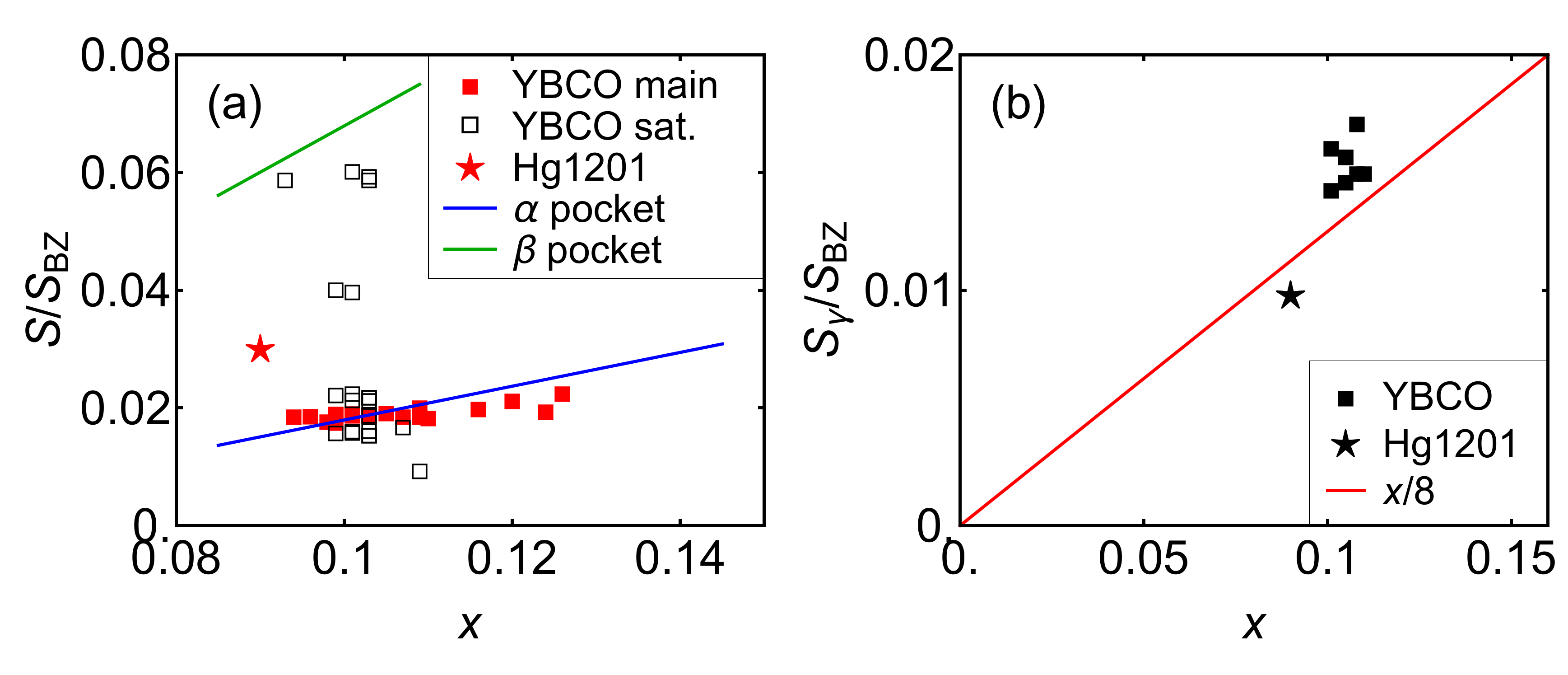}
\caption{(a) Doping dependence of the reconstructed Fermi pocket areas from DoS oscillation frequencies compared with the main and the satellite QO peaks in experiments \cite{Doiron-Leyraud2007, Jaudet2008, Sebastian2008, Bangura2008, Yelland2008, Audouard2009, Ramshaw2010, Sebastian2010b, Singleton2010, Sebastian2010, Riggs2011, Laliberte2011, Sebastian2011, Sebastian2011a, Sebastian2012a, Vignolle2013, Sebastian2014, Doiron-Leyraud2014}. (b) The area of the subdominant QO peak identified with the $\gamma$ pocket vs. doping in YBCO \cite{Audouard2009, Singleton2010, Sebastian2010b, Ramshaw2010, Vignolle2013, Sebastian2011, Sebastian2012a, Sebastian2014, Doiron-Leyraud2014} and Hg1201 \cite{Barisic2013} compared with (\ref{EqSgamma}).}
\label{FigFreq}
\end{figure}

In the small-pocket scenario, the three-peak structure of the QO spectrum can reveal rich information about the underlying pseudogap state. The dominant peak comes from the electron-like $\alpha$ pocket due to the large spectral weight on the Fermi ``arcs''. Its area versus doping quantitatively agrees with the main QO peak in YBCO as shown in Fig. \ref{FigFreq} (a). The areas of the $\alpha$ and the $\beta$ pockets depend on the CDW wavevectors thus are material-specific, but their difference satisfies the generalized Luttinger theorem: $2(S_{\beta}-S_{\alpha})=xS_{\mathrm{BZ}}$. Moreover, based on Ong's geometric construction \cite{Ong1991} and assuming isotropic scattering time we find that at relatively strong CDW order the total Hall conductance is dominated by $\alpha$ pocket and becomes negative due to its larger Fermi velocity, consistent with experiments \cite{LeBoeuf2007, LeBoeuf2011, Laliberte2011, Doiron-Leyraud2013, Badoux2015}. On the other hand, the $\gamma$ orbit area satisfies Eq. (\ref{EqSgamma}), which generally holds irrespective of materials. We extract the subdominant $\gamma$ peak frequencies from the reported QO spectra of YBCO and Hg1201 and plot them versus doping in Fig. \ref{FigFreq} (b) and find fairly good agreement with Eq. (\ref{EqSgamma}).

\section{Conclusion}
By introducing the biaxial CDW order, we show that both the bilayer model with a large Fermi surface and the pseudogap state model with small Fermi pockets can reproduce the main properties of the quantum oscillation observed in underdoped cuprates, including the multiple peaks in the oscillation spectrum and the enhancement of $m_{\alpha}^{*}$ near the CDW critical points. However, in different scenarios the subdominant $\gamma$ peak is traced back to different origins. The evolution trends of its effective mass with the CDW order strength (by tuning the doping in experiments) are qualitatively different. Therefore the quantum oscillation is a feasible tool to judge these scenarios and thus to diagnose the nature of the pseudogap state.

\acknowledgements
We are grateful to S.-S. Lee, T. M. Rice, S. Sachdev, Z.-Y. Weng  and G. Baskaran for helpful discussions. L.Z. was supported by the National Basic Research Program of China (2010CB923003). Research at Perimeter Institute is supported by the Government of Canada through Industry Canada and by the Province of Ontario through the Ministry of Research.

\bibliography{/Dropbox/ResearchNotes/BibTex/library}{}
\bibliographystyle{eplbib}
\end{document}